\documentclass[fleqn,10pt]{wlscirep}

\usepackage{bm}
\usepackage{graphicx}
\usepackage{cite}
\usepackage{color}
\usepackage{amssymb}
\usepackage{bbold}

\usepackage{xcolor} %color

\title{Tuning the Fermi velocity in Dirac materials with an electric field}

\author[1,2,*]{A. D\'{\i}az-Fern\'{a}ndez}
\author[3,4]{Leonor Chico}
\author[4,5]{J. W. Gonz\'{a}lez}
\author[1,2]{F. Dom\'{i}nguez-Adame}

\affil[1]{GISC, Departamento de F\'{\i}sica de Materiales, Universidad Complutense, E--28040 Madrid, Spain}
\affil[2]{Department of Physics, University of Warwick, Coventry, CV4 7AL, United Kingdom}
\affil[3]{Instituto de Ciencia de Materiales de Madrid, Consejo Superior de Investigaciones Cient\'{\i}ficas, C/ Sor Juana In\'{e}s de la Cruz 3, E--28049 Madrid, Spain}
\affil[4]{Donostia International Physics Center, Paseo Manuel de Lardizabal 4, E--20018 Donostia--San Sebasti\'{a}n, Spain}
\affil[5]{Centro  de  F\'{\i}sica  de  Materiales  (CSIC-UPV/EHU)--Material  Physics  Center (MPC), Paseo Manuel de Lardizabal 5, E--20018 Donostia--San Sebasti\'{a}n, Spain}

\affil[*]{alvarodi@ucm.es}

\begin{abstract}

Dirac materials are characterized by energy-momentum relations that resemble those of relativistic massless particles. Commonly denominated Dirac cones, these dispersion relations are considered to be their essential feature. These materials comprise quite diverse examples, such as graphene and topological insulators. Band-engineering techniques should aim to a full control of the parameter that characterizes the Dirac cones: the Fermi velocity. We propose a general mechanism that enables the fine-tuning of the Fermi velocity in Dirac materials in a readily accessible way for experiments. By embedding the sample in a uniform electric field, the Fermi velocity is substantially modified. We first prove this result analytically, for the surface states of a topological insulator/semiconductor interface, and postulate its universality in other Dirac materials. Then we check its correctness in carbon-based Dirac materials, namely graphene nanoribbons and nanotubes, thus showing the validity of our hypothesis in different Dirac systems by means of continuum, tight-binding and \emph{ab-initio\/} calculations.  

\end{abstract}

\begin{document}

\flushbottom

\maketitle
  
\thispagestyle{empty}

%%%%%%%%%%%%%%%%%%%%%%%%%%%%%%%%%%%%%%%%%%%%%%%%%%%%%%%%%%%
\section*{Introduction}   \label{sec:intro}
%%%%%%%%%%%%%%%%%%%%%%%%%%%%%%%%%%%%%%%%%%%%%%%%%%%%%%%%%%%

 A plethora of distinct materials are currently studied under the common feature of their low-energy excitations, resembling those of massless Dirac fermions. In addition to graphene~\cite{Novoselov04,Castro09}, two and three dimensional topological insulators and topological crystalline insulators stand out  as important examples of the so-called Dirac materials (see Ref.~\cite{Wehling14} and references therein). Kane and Mele~\cite{Kane05}, based on seminal works by Thouless \emph{et al.}~\cite{Thouless82} and Haldane~\cite{Haldane88} on the quantum Hall effect and its relation with topology, were the first to suggest in 2005 the existence of these new topological phases of matter. In 2006, Bernevig \emph{et al.}~\cite{Bernevig06} propounded a topologically insulating system, namely a HgTe/CdTe quantum well. Their prediction led to the detection in 2007 of non-trivial helical edge states, thus establishing the existence of the quantum spin Hall effect~\cite{Konig07}.

From the standpoint of applications, Dirac materials are foreseen to be of paramount importance due to their universal behaviour and the robustness of their properties, ultimately linked to symmetry~\cite{Wehling14}. Their band structure resembles the energy-momentum relation of relativistic massless particles where the energy dependence on the momentum is linear, hence the name of Dirac cones. Substantial effort is being devoted to control the slope of these cones, that is, the Fermi velocity. This parameter is essential for applications and it fully characterizes the Dirac cones.  Fermi velocity modification has been predicted and observed in few-layer graphene due to the rotation of two neighboring layers~\cite{Li2010,Trambly2010}. Indeed, in twisted bilayer graphene, a change in the band velocity was found for low rotation angles~\cite{Trambly2010}. This velocity reduction can be traced back to the interaction of the two Dirac cones, brought together due to band folding~\cite{Hicks2011}. Additionally, other mechanisms have also been put forward to produce a velocity change. For instance, Hwang \emph{et al.} demonstrated that the Fermi velocity is inversely proportional to the dielectric constant when the environment embedding graphene is modified~\cite{Hwang2012}. Many-body effects can also affect the Fermi velocity. For example, due to the divergence of the screening length at the neutrality point in graphene, a renormalization of the Fermi velocity has been related to many-body effects~\cite{Elias2011}. This many-body effect has also been detected in a topological insulator, namely, Bi$_2$Te$_3$~\cite{Miao2013}. 

Thus far, however, all these mechanisms apropos tailoring the Fermi velocity cannot be amply tuned at will, with the exception of changing the rotation angle in bilayer graphene, which implies structural modifications that are cumbersome in experiments. The aim of our work is to introduce a new mechanism, experimentally convenient, that enables fine-tuning of the Fermi velocity. We shall prove that double-gated Dirac materials are excellent candidates to verify this effect. 

Firstly, we predict a velocity reduction in a topological insulator/semiconductor (TI/S) junction where a uniform electric field is applied perpendicular to the interface. Surface states lying within the gap are known to occur and their description can be made in terms of a Dirac-like equation where the bandgap changes sign across the interface (see Refs.~\cite{Volkov85,Korenman87,Agassi88,Pankratov90,Kolesnikov97} and references therein), similar to the Jackiw and Rebbi model for solitons~\cite{Jackiw76}. We have found that these surface states display a Dirac cone dispersion that widens as the field increases. The exact solution, although feasible, is rather intricate so it is left for the Supplemental Information. Instead, an approximate solution is presented and an analytical expression displaying the renormalization of the Fermi velocity is obtained.   

 Secondly, bearing in mind the universal properties of Dirac materials, we verify the generality of our results in other Dirac systems, such as graphene nanoribbons (GNRs) and carbon nanotubes (CNTs), and employing a different method, namely, the tight-binding (TB) approximation. We show that both TB and low-energy continuum (Dirac equation) calculations confirm the possibility to adjust the Fermi velocity in Dirac materials by lowering the slope of the Dirac dispersion relation. This trend is also confirmed by \emph{ab-initio\/} calculations. Hence, we conclude that the mechanism is completely valid for systems with a linear dispersion relation. Experimental proposals towards the detection of this phenomenon are discussed at the end of the paper.

%%%%%%%%%%%%%%%%%%%%%%%%%%%%%%%%%%%%%%%%%%%%%%%%%%%%%%%%%%%
\section*{Results}   \label{sec:results}
%%%%%%%%%%%%%%%%%%%%%%%%%%%%%%%%%%%%%%%%%%%%%%%%%%%%%%%%%%%

%%%%%%%%%%%%%%%%%%%%%%%%%%%%%%%%%%%%%%%%%%%%%%%%%%%%%%%%%%%
\subsection*{TI/S junction under bias: continuum model}   \label{sec:single}
%%%%%%%%%%%%%%%%%%%%%%%%%%%%%%%%%%%%%%%%%%%%%%%%%%%%%%%%%%%

Throughout this section, we will focus on IV-VI semiconductors, among which the ternary compounds Pb$_{1-x}$Sn$_{x}$Te and Pb$_{1-x}$Sn$_{x}$Se can be found. The latter are known to shift from being semiconductors to topological crystalline insulators due to the band inversion that occurs at the $L$ points of the Brillouin zone as the fraction of Sn increases~\cite{Assaf2016}. Hence, a TI/S junction can be built using these materials.

%%%%%%%%%%%%%%%%%%%%%%%%%%%%%%%%%%%%%%%%%%%%%%%%%%%%%%%%%%%
%\subsubsection*{Unbiased junction}   \label{sec:unbiased}
%%%%%%%%%%%%%%%%%%%%%%%%%%%%%%%%%%%%%%%%%%%%%%%%%%%%%%%%%%%

The envelope function calculation is a reliable method to obtain the electron states near the band edges in narrow-gap IV-VI semiconductors~\cite{Kriechbaum86}.  
Keeping only two nearby bands and neglecting the influence of far bands, electron states near the $L$ extrema are determined from the following Dirac-like Hamiltonian~\cite{Agassi88,Pankratov90} 
\begin{equation}
\mathcal{H}_{0}=v_{\bot}{\bm\alpha}_{\bot}\cdot{\bm p}_{\bot}+v_z\alpha_z p_z
+\frac{1}{2}\,E_{\mathrm{G}}(z)\beta+V_{\mathrm C}(z)\ ,
\label{eq:01}
\end{equation}
where the $Z$ axis is parallel to the growth direction $[111]$. It is understood that the subscript $\bot$ of a vector indicates the nullification of its $z$-component. Here $E_{\mathrm{G}}(z)$ stands for the position-dependent gap and $V_{\mathrm C}(z)$ gives the position of the gap centre. ${\bm\alpha}=(\alpha_x,\alpha_y,\alpha_z)$ and $\beta$ denote the usual $4\times 4$ Dirac matrices, $\alpha_{i}=\sigma_x\otimes \sigma_{i}$ and $\beta=\sigma_z\otimes\mathbb{1}_2$, $\sigma_i$ and $\mathbb{1}_n$ being the Pauli matrices and $n\times n$ identity matrix, respectively. Here $v_{\bot}$ and $v_z$ are interband matrix elements having dimensions of velocity. Assuming that the surface states spread over distances along the growth direction much larger than the interface region, we can confidently consider an abrupt profile for both the magnitude of the gap $E_{\mathrm{G}}(z)=2\Delta_{\mathrm{L}}\theta(-z)+2\Delta_{\mathrm{R}}\theta(z)$ and the gap centre $V_{\mathrm{C}}(z)=V_{\mathrm{L}}\theta(-z)+V_{\mathrm{R}}\theta(z)$, where $\theta(z)$ is the Heaviside step function. The subscripts L and R refer to the left and right sides of the heterojunction, respectively. Note that in the case of a TI/S junction $\Delta_{\mathrm{L}}\Delta_{\mathrm{R}}<0$.

The Hamiltonian~(\ref{eq:01}) acts upon the envelope function ${\bm\chi}({\bm r})$, which is a four-component vector involving the two-component spinors ${\bm\chi}_{+}({\bm r})$ and ${\bm\chi}_{-}({\bm r})$ belonging to the $L^{+}$ and $L^{-}$ bands. Since the interface momentum is conserved, the envelope function can be written as ${\bm\chi}({\bm r})={\bm\chi}(z)\exp(i{\bm r}_{\bot}\cdot{\bm k}_{\bot})$, where ${\bm\chi}(z)$ decays exponentially on each semiconductor. In the case of symmetric and same-sized gaps, namely $V_{\mathrm{C}}(z)=0$ and $E_{\mathrm{G}}(z)=2\Delta\,\mathrm{sgn}(z)$, it is found that ${\bm\chi}(z)\sim \exp(-|z|/d)$, with $d=\hbar v_{\bot}/\Delta$ and the interface dispersion relation is a single Dirac cone $E({\bm k}_{\bot})=\pm \hbar v_{\mathrm{F}}|{\bm k}_{\bot}|$, where $v_{\mathrm{F}}=v_{\bot}$ (see Ref.~\cite{Adame94} for further details).

%%%%%%%%%%%%%%%%%%%%%%%%%%%%%%%%%%%%%%%%%%%%%%%%%%%%%%%%%%%
%\subsubsection*{Biased junction}   \label{sec:biased}
%%%%%%%%%%%%%%%%%%%%%%%%%%%%%%%%%%%%%%%%%%%%%%%%%%%%%%%%%%%

We now turn to the electronic states of a single TI/S junction subjected to a perpendicular electric field ${\bm F}=F\,\widehat{\bm z}$. Notice that ${\bm k}_{\bot}$ is still conserved. The calculation of the states in the quantum-confined Stark regime for general values of the parameters $\Delta_{\mathrm{L,R}}$ and $V_{\mathrm{L,R}}$ is feasible but rather involved. In order to keep the algebra as simple as possible, we restrict ourselves to symmetric and same-sized gaps. This is not a serious limitation but the calculations are considerably simplified. 

The electrostatic potential is $-eFz$ when the electric field is switched on and the corresponding Dirac equation  becomes $(\mathcal{H}_0-eFz-E){\bm\chi}(z)=0$. Although this problem is exactly solvable (see Supplemental Information for details), we shall follow a more ingenious approach by introducing the Feynman-Gell-Mann \emph{ansatz}, ${\bm\chi}(z)=(\mathcal{H}_0+eFz+E){\bm\psi}(z)$~\cite{Feynman58}. Defining the following dimensionless quantities ${\bm\kappa}={\bm k}_{\bot}d$, $\xi=z/d$, $\varepsilon=E/\Delta$, $f=eFd/\Delta$, we obtain
\begin{equation}
\Big[-\frac{d^2}{d\xi^2}+U(\xi)-f^2\xi^2
- 2\varepsilon f\xi-if\alpha_z+\lambda^{2}\Big]{\bm \psi}(\xi)=0\ ,
\label{eq:07}
\end{equation}
where $U(\xi)=2i\beta\alpha_z\delta(\xi)$ and
\begin{equation}
\lambda^{2}=\kappa^2+1-\varepsilon^2\ .
\label{eq:08}
\end{equation}
Notice that we restrict ourselves to isotropic junctions ($v_z=v_{\bot}$) for simplicity, an assumption that can easily be relaxed.

In order to further simplify the problem, we will focus on electric fields smaller than $F_\mathrm{C}\equiv \Delta/ed=\Delta^2/e\hbar v_{\bot}$. In other words, we will assume that the potential energy drop across the distance $d$ over which the surface states spread along the growth direction is smaller than half of the gap $\Delta$. It should be stressed that this is the usual regime in experiments. Typical values for IV-VI compounds are $\Delta=75\,$meV and $d=4.5\,$nm~\cite{Korenman87}, yielding $F_\mathrm{C}=170\,$kV/cm, which is actually large. Thus, under the reasonable assumption that $F < F_{\mathrm{C}}$ one can neglect the quadratic term $f^2\xi^2$ in equation~(\ref{eq:07}). The constant matrix term $-if\alpha_z$ is easily diagonalized by a unitary transformation. Nevertheless, we have checked that it has a negligible impact on the final results even at moderate fields and it will be omitted as well (see Supplemental Information for a comparison between the exact calculations and the approximate ones reported here). 
 
We can regard the term $U(\xi)$ in~(\ref{eq:07}) as a perturbation and seek for the retarded Green's function of the unperturbed problem. It can be cast in the form $\mathcal{G}_{0}^{+}(\xi,\xi^{\prime};\varepsilon)=G_{0}^{+}(\xi,\xi^{\prime};\varepsilon)\mathbb{1}_4$, where the scalar Green's function obeys the following equation
\begin{equation}
\left[-\frac{\partial^2}{\partial\xi^2}-2\varepsilon f\xi 
+\lambda^{2}\right]G_{0}^{+}(\xi,\xi^{\prime};\varepsilon)=
\delta(\xi-\xi^{\prime})\ .
\label{eq:11}
\end{equation}
where it is understood that $\mathrm{Im}(\lambda^2)<0$. Since we are interested in states confined along the growth direction, we consider $\mathrm{Re}(\lambda^2)>0$ in what follows. 

Equation~(\ref{eq:11}) is analogous to the problem of a non-relativistic one-dimensional particle in a tilted potential solved in Refs.~\cite{Ludviksson87,Jung09}. Let us define
\begin{equation}
\mu=(2|\varepsilon| f)^{1/3}\ ,
\quad
p(\xi)=-s_{\varepsilon}\mu\xi+\frac{\lambda^{2}}{\mu^{2}}\ ,
\label{eq:12}
\end{equation}
with the shorthand notation $s_{\varepsilon}=\mathrm{sgn}\left[\mathrm{Re}(\varepsilon)\right]$. In terms of these parameters the retarded Green's function for the appropriate boundary conditions is
\begin{eqnarray}
G_{0}^{+}(\xi,\xi^{\prime};\varepsilon)&=&-\frac{\pi s_{\varepsilon}}{\mu}\Big\{
\theta\left[(\xi^{\prime}-\xi)s_{\varepsilon}\right]
\mathrm{Ai}\left(p(\xi)\right)
\mathrm{Ci}^{+}\left(p(\xi^{\prime})\right)
\nonumber\\
&+&
\theta\left[(\xi-\xi^{\prime})s_{\varepsilon}\right]\mathrm{Ai}\left(p(\xi^{\prime})\right)
\mathrm{Ci}^{+}\left(p(\xi)\right)\Big\}\ ,
\label{eq:13}
\end{eqnarray}
where $\mathrm{Ci}^{+}(z)=\mathrm{Bi}(z)+i\mathrm{Ai}(z)$, $\mathrm{Ai}(z)$ and $\mathrm{Bi}(z)$ being the Airy functions~\cite{Abramowitz72}. 

Once the retarded Green's function for the unperturbed problem is known, the Dyson equation is exactly solved to obtain the Green's function $\mathcal{G}^{+}(\xi,\xi^{\prime};\varepsilon)$ associated to Eq.~(\ref{eq:07}). This function is analytic in the lower half plane $\mathrm{Im}(\lambda^2)<0$. Thus, it may have simple poles when it is analytically continued to the upper half plane~\cite{Jung09}. In our case, the poles satisfy the condition
$\left[G_0^{+}(0,0;\varepsilon)\right]^{2}=1/4$. From~(\ref{eq:13}) we finally get 
\begin{equation}
\frac{4\pi^2}{\mu^{2}}\left[\mathrm{Ai}\left(\frac{\lambda^{2}}{\mu^{2}}\right)\mathrm{Ci}^{+}\left(\frac{\lambda^{2}}{\mu^{2}}\right)\right]^2=1\ .
\label{eq:16}
\end{equation}

Since the complex poles are found for $\mathrm{Im}(\lambda^2)>0$, we must then have $\mathrm{Re}(\varepsilon)\mathrm{Im}(\varepsilon)<0$. That is, we may write the energy as $\varepsilon_{\pm}=\varepsilon_r\pm i\gamma/2$, where the plus (minus) sign is to be considered when $\varepsilon_r$ is negative (positive). Here $\gamma$ is a real positive number. Hence, considering the aforementioned prescription for the relative signs of the real and imaginary parts of the poles, we can find these by solving~(\ref{eq:16}). Poles correspond to resonant states with energy $E({\bm \kappa}_{\bot})=\varepsilon_r\Delta$ and level width $\Gamma({\bm \kappa}_{\bot})=\gamma\Delta$. Therefore, in the presence of the electric field, electrons can tunnel into the continuum and escape from the TI/S junction due to the non-zero imaginary part of the pole. This is a common feature in the quantum-confined Stark effect~\cite{Miller84}. Nonetheless, in the next section we will show that the level width is exponentially small in the low-field regime, that is, tunnelling is only important at high fields.

%%%%%%%%%%%%%%%%%%%%%%%%%%%%%%%%%%%%%%%%%%%%%%%%%%%%%%%%%%%
%\subsubsection*{Low-field limit}   \label{sec:low-field}
%%%%%%%%%%%%%%%%%%%%%%%%%%%%%%%%%%%%%%%%%%%%%%%%%%%%%%%%%%%

We can simplify~(\ref{eq:16}) in the low-field regime $F<F_\mathrm{C}$ by noticing that $|\lambda|\gg \mu$. In this limiting case we approximate the Airy functions to their asymptotic expansions for large argument and obtain (see the Supplemental Information for details)
\begin{subequations}
\begin{eqnarray}
E({\bm k}_{\bot})&=&\pm \hbar v_{\bot}
\left(1-\frac{5}{8}\frac{F^2}{F^{2}_{\mathrm{C}}}\right)|{\bm k}_{\bot}|\ ,
\label{eq:17a}\\
 \Gamma({\bm k}_{\bot})&=&\frac{\Delta^2}{\hbar v_{\bot}|{\bm k}_{\bot}|}\,
\exp\left(-\frac{2\Delta}{3\hbar v_{\bot}|{\bm k}_{\bot}|}\,
\frac{F_\mathrm{C}}{F}\right)\ .
\label{eq:17b}
\end{eqnarray}
\end{subequations}

Equation~(\ref{eq:17a}) is very remarkable and it is our main result. It means that applying an electric field perpendicular to the junction the interface linear dispersion remains but we are effectively lowering the Fermi velocity
\begin{equation}
v_F(F)\equiv  v_{\bot}\left(1-\frac{5}{8}\frac{F^2}{F^{2}_{\mathrm{C}}}\right)\ .
\label{eq:18}
\end{equation}
We still get a cone but it becomes wider. In addition, as it was anticipated above, equation~(\ref{eq:17b}) ensures that the tunnelling rate into the continuum is exponentially small in the low-field limit.

\begin{figure}[htb]
\centerline{\includegraphics[width=0.95\columnwidth]{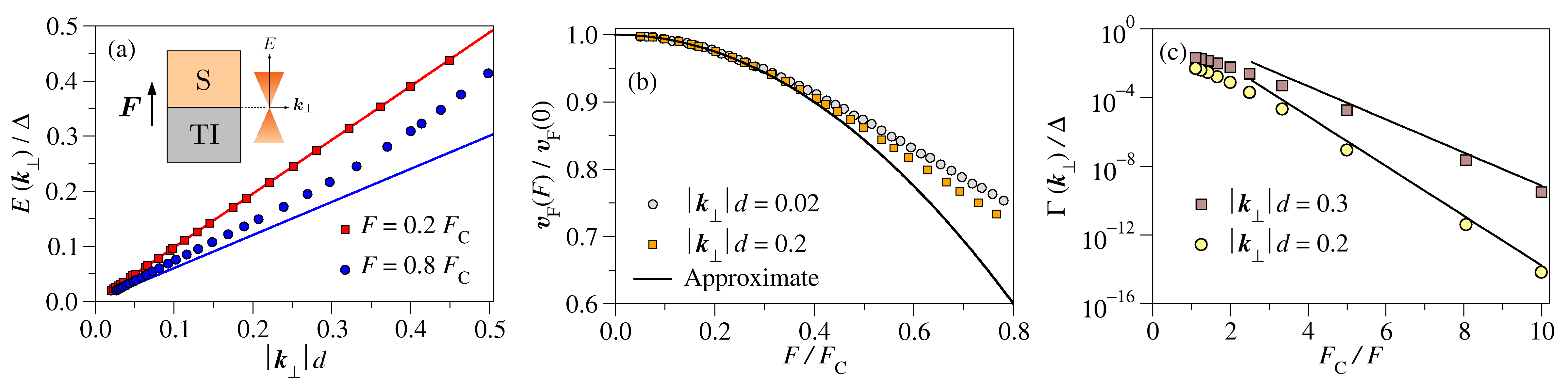}}
\caption{(a)~Energy as a function of the interface momentum for two values of the electric field. Solid lines correspond to the approximate result~(\ref{eq:17a}). The dispersion grows faster than linear only at high field. (b)~Fermi velocity as a function of the the electric field for two different values of the interface momentum. Solid line displays the quantum-confined Stark effect predicted by~(\ref{eq:18}). (c)~Level width as a function of the inverse of the electric field for $|{\bm k}_{\bot}|d=0.2$ (circles) and $|{\bm k}_{\bot}|d=0.3$ (squares). Solid lines depict the approximation given in~(\ref{eq:17b}).}
\label{fig1}
\end{figure}

In order to verify results~(\ref{eq:17a}) and~(\ref{eq:17b}), we numerically tested them from the exact numerical solution of equation~(\ref{eq:16}). Figure~\ref{fig1}(a) shows the real part of the pole of $\mathcal{G}^{+}(\xi,\xi^{\prime};\varepsilon)$ as well as the approximate expression~(\ref{eq:17a}) as a function of the interface momentum. At low electric field ($F=0.2F_{\mathrm{C}}$) the dispersion is perfectly linear and the accuracy of our approximation~(\ref{eq:17a}) is outstanding. Slight deviations from the linear behaviour appear upon increasing the electric field ($F=0.8F_{\mathrm{C}}$) and the approximate slope is no longer valid, as expected. The maximum value of the electric field below which the  approximation~(\ref{eq:17a}) holds can be estimated from figure~\ref{fig1}(b). It displays the Fermi velocity as a function of the electric field for two different values of the interface momentum. It is seen that the approximate result~(\ref{eq:18}) fits the exact result in the range $F \lesssim 0.4 F_{\mathrm{C}}$. Figure~\ref{fig1}(c) displays the level width obtained from the imaginary part of the pole of $\mathcal{G}^{+}(\xi,\xi^{\prime};\varepsilon)$ as a function of the inverse of the electric field for two different values of $|{\bm k}_{\bot}|d$. Solid lines show the approximate width given in~(\ref{eq:17b}). There exists a very good agreement with exact results at moderate fields ($F \lesssim 0.4 F_{\mathrm{C}}$). It is worth mentioning that the exact level width is much lower than the approximate value at $F \simeq F_{\mathrm{C}}$. Thus, we conclude that tunnelling is not important except at high fields.

\begin{figure}[tb]
\centerline{\includegraphics[width=0.3\columnwidth]{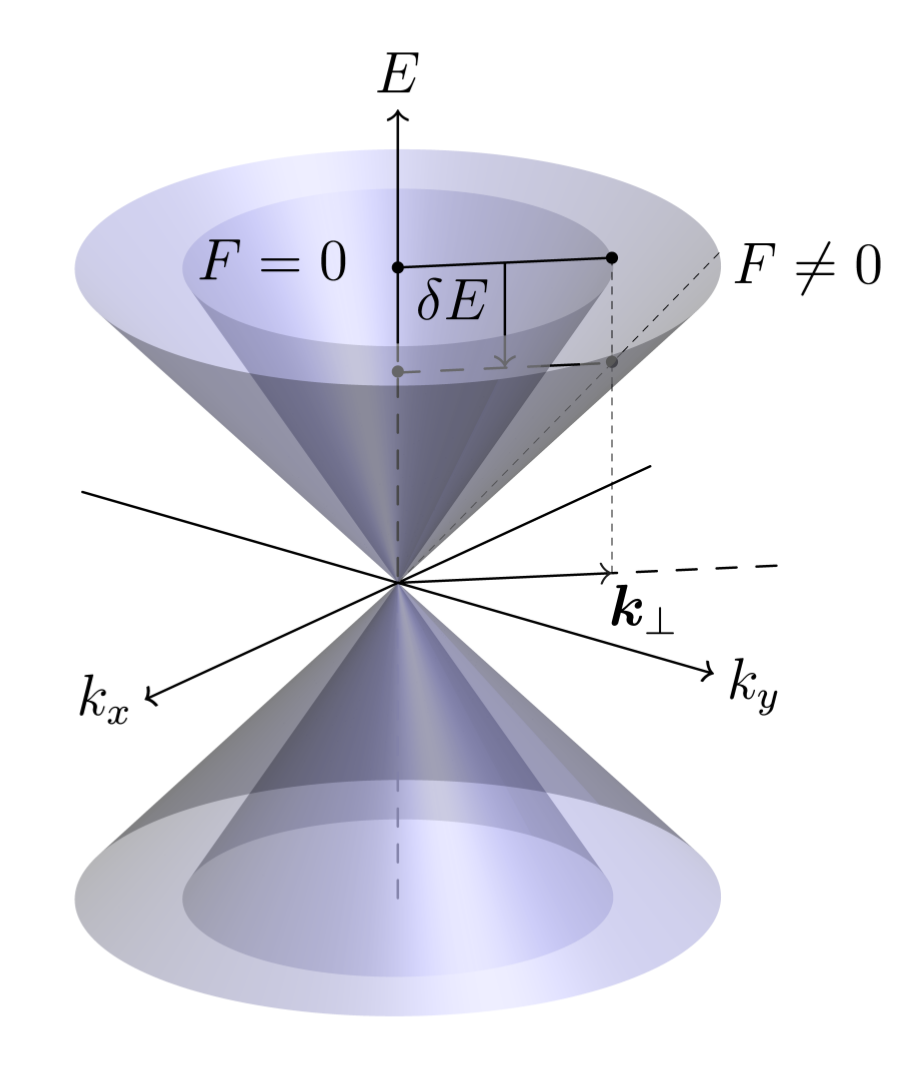}}
\caption{Dirac cones of the surface states at zero and finite electric field ${\bm F}$, applied perpendicular to the interface. When the electric field is adiabatically switched on, the magnitude of the electron energy decreases by an amount $\delta E$ and the conservation of the interface momentum, $\bm{k}_{\perp}$, gives rise to a widened cone.}
\label{fig2}
\end{figure}

A qualitative explanation of the reduction of the Fermi velocity can be traced back to the quantum-confined Stark effect. For concreteness, let us focus on surface states with positive energy. Before the electric field is applied, surface states are exponentially localized at the junction, with gapless and linear dispersion $\hbar v_{\bot}|{\bm k}_{\bot}|$. When the electric field is adiabatically applied, it is well known that perturbation theory establishes that the energy is quadratically lowered with the field by an amount $\delta E$ (see figure~\ref{fig2}). But the interface momentum is conserved if the field is perpendicular to the TI/S junction. Therefore, the Dirac cone widens, as schematically shown in figure~\ref{fig2}, and the Fermi velocity is in effect quadratically lowered with the electric field. It should be stressed that equation~(\ref{eq:11}) becomes independent of the electric field if $\varepsilon=0$. Consequently $\varepsilon=0$ for ${\bm \kappa}_{\bot}=0$ is still an eigenenergy of the system when the electric field is applied and the dispersion remains gapless. In other words, the magnitude of the energy shift $\delta E$ due to the electric field must decrease upon decreasing the energy.

In view of the above results, we consider that the Fermi velocity reduction is related to the existence of a Dirac cone, i.e., a linear dispersion relation, so it should be achievable in other Dirac materials. We choose carbon-based systems, such as GNRs and CNTs, as the easiest examples in which to validate this hypothesis with a simple TB model, which nonetheless includes other bands not present in the continuum approach presented before. Next section will be devoted to confirm this idea by applying a uniform electric field across a metallic armchair graphene nanoribbon (aGNR), a metallic armchair carbon nanotube (aCNT) and a metallic zigzag carbon nanotube (zCNT).

%%%%%%%%%%%%%%%%%%%%%%%%%%%%%%%%%%%%%%%%%%%%%%%%%%%%%%%%%%%
\subsection*{Carbon-based Dirac materials: tight-binding and ab-initio calculations}  \label{sec:carbon}
%%%%%%%%%%%%%%%%%%%%%%%%%%%%%%%%%%%%%%%%%%%%%%%%%%%%%%%%%%%

\begin{figure}[tb]
\centerline{\includegraphics[width=0.95\columnwidth]{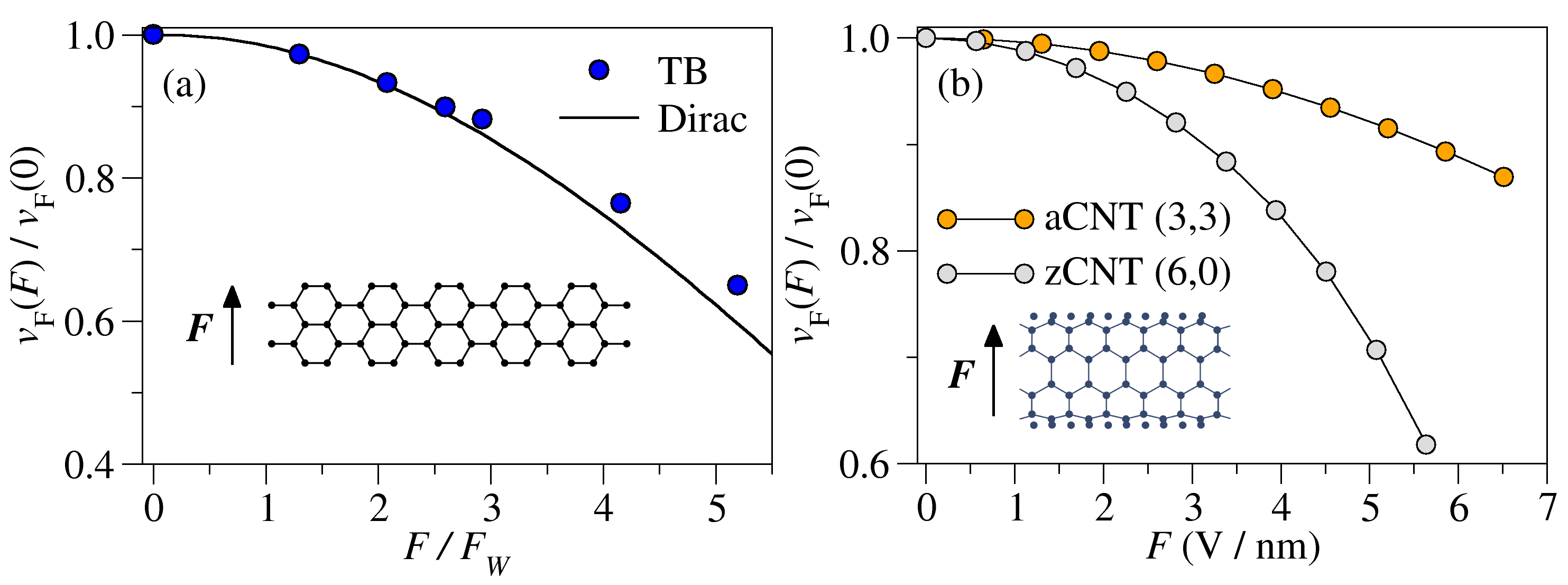}}
\caption{(a)~Fermi velocity for a metallic aGNR as a function of the normalized field $F/F_{\text{W}}$ (see main text for details). A prominent agreement between the TB approach and the Dirac equation is observed. (b)~Fermi velocity for a metallic aCNT and a zCNT as a function of the field $F=\Delta V/d$, where $\Delta V$ is the potential drop across the CNT and $d$ is its diameter. $v_F(F=0)=8.617\times 10^5\,$m/s for the zCNT and $v_F(F=0)=8.010\times 10^5\,$m/s for the aCNT. In all cases, $v_F$ is significantly reduced when the field is switched on.}
\label{fig3}
\end{figure}

In order to elucidate the generality of our result, we check for its occurrence in carbon-based Dirac materials, such as metallic aGNRs, aCNTs and zCNTs. These systems can be easily described with a one-orbital nearest-neighbour hopping tight-binding approximation with an electric field term. TB calculations confirm that the Fermi velocity is remarkably reduced, and the results for the aGNR are further validated by means of a low-energy Dirac equation. Our results are summarized in figure~\ref{fig3}, leaving the details of the calculations for the Supplemental Information.
As can be immediately noticed, all systems display a significant reduction of the Fermi velocity as a function of the electric field. Our conjecture is then firmly established: the Fermi velocity can be reduced via a uniform electric field. In the case of a metallic aGNR, shown in figure~\ref{fig3}(a), the agreement between the TB and Dirac approaches is noteworthy. It is worth mentioning that the approaches start to slightly differ when the normalized field $f=F/F_{W}$ is $f\gtrsim 3$, being $F_{W}=\hbar v_{F}(0)/e(W+a)^2$, where $W$ is the width of the aGNR, $a=0.246\,$nm, and $v_{F}(0)$ is the Fermi velocity of unbiased graphene. Actually, a small gap opens up in the TB calculations due to the interaction with other bands. This gap is negligible in the region where the Dirac and TB approaches coincide. However, as it can be drawn from figure~\ref{fig3}(a), these fields are remarkably large even from a practical perspective. For instance, an aGNR with $W\sim 2\,$nm has $F_W\sim 0.1\,$V/nm, meaning that the system remains effectively gapless even for very large applied fields of the order of a few tenths of V/nm. Boundary conditions ensuring the nanoribbon to be metallic~\cite{Wurm11} are preserved even in the presence of the electric field, thus not opening a gap for any value of $F$ when the continuum description is considered. This is not the case for zGNRs, where TB calculations show that any strength of the electric field opens a sizable gap. Nevertheless, it is well-known that zGNRs do not show linear dispersion relations. In fact, for all the cases in which the material has a linear dispersion relation, i.e., a Dirac cone-shaped band, we find a similar behaviour, with a noticeable Fermi velocity reduction.  

\begin{figure}[htb]
\centerline{\includegraphics[width=0.6\columnwidth]{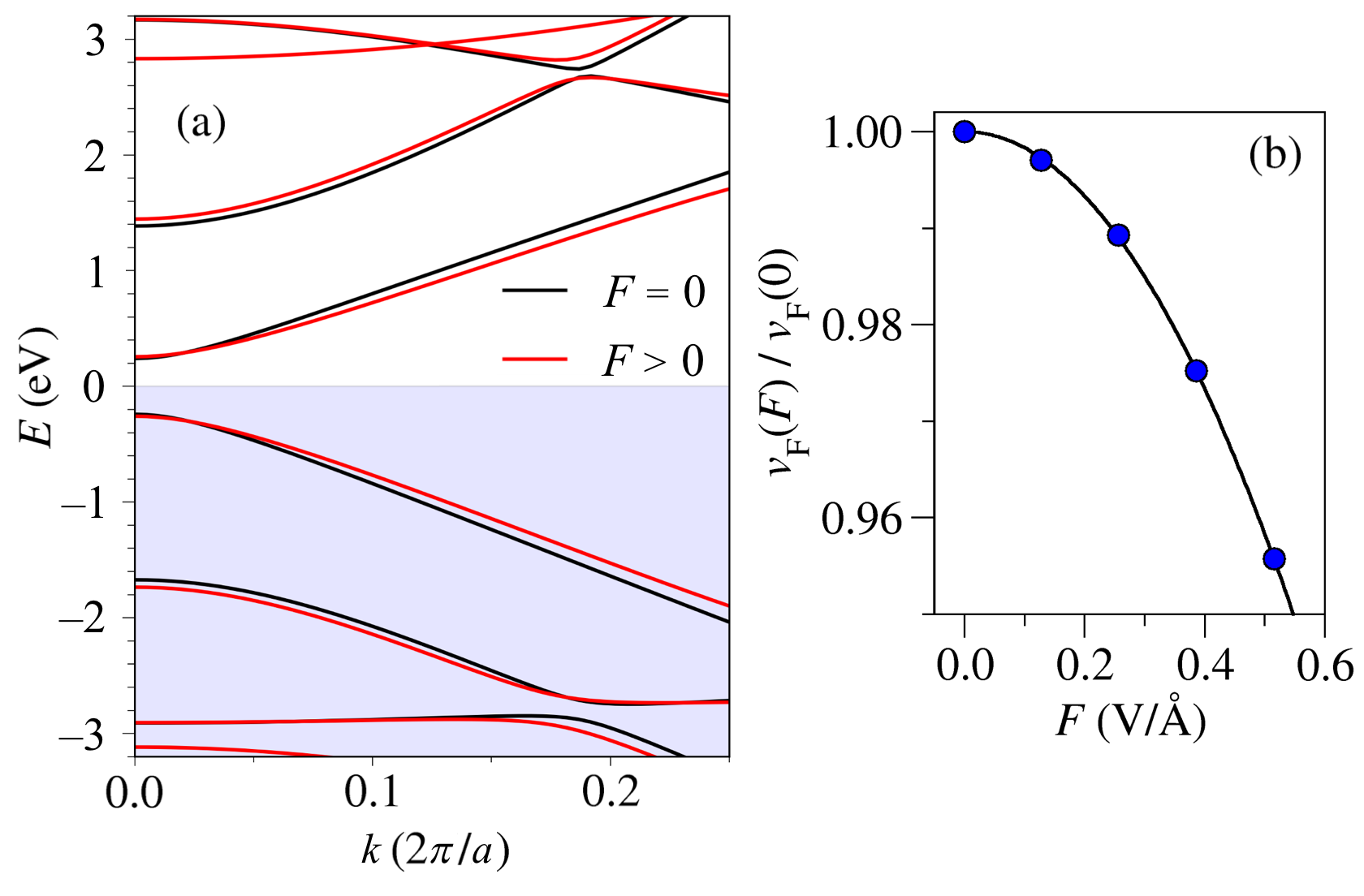}}
\caption{
(a)~DFT-calculated band structure of an $N=5$ aGNR with~(red) and without~(black) a transversal electric field $F=0.51\,$V/\AA. (b)~Variation of the Fermi velocity of the first conduction band as a function of the electric field. We have considered the slope for the linear bands between $k\sim 0.1\times 2\pi/a$ and $k\sim 0.2\times 2\pi/a$.
}
\label{fig4}
\end{figure}

We have also verified our results by means of \emph{ab-initio\/} calculations based on the density functional theory~(DFT). We explore the effect of the applied electric field on an $N=5$ aGNR, with $N=2M+1$, $M$ being the number of hexagons across the nanoribbon. It turns out that this nanoribbon is near-metallic within this approach\cite{kimouche2015ultra}. Figure~\ref{fig4}(a) shows the DFT-calculated band structures with and without an external field of strength $F=0.51\,$V/\AA. There is a clear reduction of the velocity in the linear parts of the valence and conduction bands closest to $E_F$, thus corroborating the observed tunability of the Fermi velocity with a more sophisticated approach. Figure~\ref{fig4}(b) depicts the variation of the Fermi velocity as a function of the electric field, which presents the same quantitative behaviour observed with the continuum and tight-binding methods. Due to electronic correlations, electric polarizability and charge screening effects, a direct comparison between the effective or semi-empirical models and those used in DFT calculations is not straightforward. Nonetheless, the predicted tendency is clearly confirmed.

%%%%%%%%%%%%%%%%%%%%%%%%%%%%%%%%%%%%%%%%%%%%%%%%%%%%%%%%%%%
\subsection*{Experimental proposals}
%%%%%%%%%%%%%%%%%%%%%%%%%%%%%%%%%%%%%%%%%%%%%%%%%%%%%%%%%%%

The most direct way to detect a variation of $v_F$ in a Dirac material is to measure the energy dispersion relation in an ARPES experiment~\cite{Miao2013}. Notice that, due to the linear dispersion near the Fermi point, the conductivity does not depend on the particular value of the band slope in an undoped system. However, in a doped sample, a Fermi velocity dependence is expected, both in the pristine and disordered case, as in graphene~\cite{DasSarma2011,Kim2012,Joucken2015}. Another way to detect the change in the Fermi velocity is to apply the electric field in a limited region and inject carriers from a field-free contact of the same system. Then the conductance should  decrease as a function of the applied field due to the wave vector mismatch~\cite{Chico2004,Gonzalez2010}. In any case, one of the most representative features of the Fermi velocity modulation is the dependence of the Landau levels on $v_F$~\cite{Assaf2016}. Thus, magnetotransport measurements will unambiguously evidence any modification of this parameter. 

The typical experimental uncertainty in the measurement of $v_F$ is about $2\%$~\cite{Assaf2016}, and this implies from~(\ref{eq:18}) that $F$ should be larger than $0.2F_{\mathrm{C}}$ to observe a change in the Fermi velocity in the TI/S junction. Nonetheless,  this value is still within the low-field range mentioned above and, consequently, the approximate Fermi velocity given in~(\ref{eq:18}) holds in experiments. Other physical properties, such as the effective fine structure constant, related to the strength of many-body interactions, will also be affected by the variation of $v_F$ and tuned by an external electric field during an experiment~\cite{Jang2008}. The reduction of the Fermi velocity is an effect with measurable consequences on several physical magnitudes, and we expect it to have applications for the design of novel devices based on Dirac materials. One possibility that deserves to be explored is the change in the electronic conductance in a Dirac material with a finite region where an electric field is applied. The wavevector mismatch due to the change of velocities may be used to control the conductance through the system, thus paving the way to devices based on this effect.  

%%%%%%%%%%%%%%%%%%%%%%%%%%%%%%%%%%%%%%%%%%%%%%%%%%%%%%%%%%%
\section*{Discussion}
%%%%%%%%%%%%%%%%%%%%%%%%%%%%%%%%%%%%%%%%%%%%%%%%%%%%%%%%%%%

Dirac materials share common characteristics due to their linear dispersions, also known as Dirac cones. A myriad of applications will benefit from their remarkable properties. However, there is an increased need to control their original properties by simple techniques so as to boost their potential applications. In this work, we propose the use of uniform electric fields to tune their electronic properties, proving that a significant Fermi velocity reduction can be efficiently achieved. Two archetypal families of Dirac systems are studied: a TI/S junction under a uniform electric field perpendicular to the interface, and carbon-based materials, such as metallic aGNRs and CNTs with an applied electric field. Analytical expressions are obtained for the TI/S junction by solving a spinful two-band model that is equivalent to the Dirac model for relativistic electrons. The mass term is half the bandgap and changes its sign across the junction. Under certain reasonable assumptions we have obtained a closed expression for the energy levels. These levels turn out to be narrow resonances upon applying the field. In  particular, it is a remarkable result that the interface linear dispersion is preserved and the Fermi velocity is simply renormalized by the electric field. We have related the lowering of the Fermi velocity to the conservation of the interface momentum after the adiabatic increase of the electric field.

Given the generality of our model, we postulate our results to be true for all Dirac materials. We have verified this in carbon-based Dirac systems such as metallic aGNRs and CNTs, where the same phenomenon is observed. Additionally, we suggest a series of feasible ways to detect the Fermi velocity reduction via spectroscopic and transport measurements. 

%%%%%%%%%%%%%%%%%%%%%%%%%%%%%%%%%%%%%%%%%%%%%%%%%%%%%%%%%%%
\section*{Methods}
%%%%%%%%%%%%%%%%%%%%%%%%%%%%%%%%%%%%%%%%%%%%%%%%%%%%%%%%%%%

The tight-binding numerical simulations were done with the tight-binding Hamiltonian on a honeycomb lattice with rectangular shape. Both the upper and lower edge are taken to be of armchair or zigzag type. The Hamiltonian can be written as $H=H_0 + H_F$, where $H_0$ is the kinetic energy term, $H_0 = -t \sum_{\langle i,j\rangle}  c_{i}^\dagger c_{j}^{}$, with $t = 2.67$ eV being the nearest-neighbour hopping energy and $c_{i}$, $c_{j}^\dagger$  the destruction and creation operators for an electron on sites $i$ and $j$, respectively. The in-plane uniform electric field is applied transverse to the nanoribbon and its contribution is described as $H_F= -e{\bm F}\cdot  \sum_i {\bm r}_i \, c_{i}^\dagger c_{i}^{}$ where ${\bm F}$ is the electric field and ${\bm r}_i$ is the position of atomic site~$i$. 

Density functional theory calculations were performed using the plane-wave self-consistent field plane-wave implemented in the {\sc Quantum ESPRESSO} package\cite{QE-2009} with the generalized gradient approximation of Perdew-Burke-Ernzerhof exchange-correlation functional\cite{perdew1996generalized}. Self-consistent charge calculations are converged up to a tolerance of $10^{-8}$ for the unit cell. Ribbons are repeated periodically separated by $15\,$\AA\ of empty space in the perpendicular directions and hydrogen atoms saturate the dangling bonds of the edge carbon atoms. A fine $k$-grid of $25 \times 1 \times 1$ Monkhorst-Pack is used to sample the Brillouin zone and the orbitals were expanded in plane waves until a kinetic energy cutoff of $815\,$eV.  All atoms are allowed to relax within the conjugate gradient method until forces have been converged with a tolerance of $10^{-3}\,$eV/\AA. The external electric field was included in the {\sc Quantum ESPRESSO} calculations using two different approaches: by adding a sawtooth potential transversal to the ribbon direction\cite{resta1986self} and through the modern theory of polarization (Berry phases)\cite{resta2007theory,souza2002first,umari2002ab}. Both approaches confirm the tunability of the Fermi velocity.

\bibliography{references}

\section*{Acknowledgements}

The authors are indebted to M.~Saiz-Bret\'{\i}n and A. Ayuela for helpful discussions. A.~D-F.\ and F.~D-A.\ thank the Theoretical Physics Group of the University of Warwick for their warm hospitality. L.~C.\ gratefully acknowledges the hospitality of the Donostia International Physics Center. This work was supported by the Spanish MINECO under grants MAT2013-46308, MAT2016-75955, FIS2015-64654-P, FIS2013-48286-C02-01-P and FIS2016-76617-P, by the Basque Government through the ELKARTEK project (SUPER) and the University of the Basque Country (Grant No. IT-756-13).

\section*{Author contributions statement}
A.D-F. and F.D-A. performed the analytical calculations based on the Dirac equation. J.W.G. performed the numerical simulations in carbon-based systems. A.D-F., L.C., J.W.G. and F.D-A. all contributed to conceptual developments and manuscript preparation.

\section*{Additional information}

\textbf{Supplementary information} accompanies this paper at http://www.nature.com/srep; 
\textbf{Competing financial interests:} The authors declare no competing financial interests.

\end{document}